\documentclass{kapproc} 

\usepackage{procps} 

\usepackage[dvips]{graphicx}

\upperandlowercase

\kluwerbib 

\begin{document}

\articletitle[First phylogenetic analyses of galaxy evolution]
{First phylogenetic analyses \\
of galaxy evolution}

\author{Didier Fraix-Burnet} 
 
\affil{Laboratoire d'Astrophysique de Grenoble
             BP 53, F-38041, Grenoble cedex 9 (France)
}

\begin{abstract}
The Hubble tuning fork diagram, based on morphology, has always been the
preferred scheme for classification of galaxies and is still the only one originally
built from historical/evolutionary relationships. At the opposite, biologists have
long taken into account the parenthood links of living entities for
classification purposes. Assuming branching evolution of galaxies as a 'descent
with modification', we show that the concepts and tools of phylogenetic
systematics widely used in biology can be heuristically transposed to the case of
galaxies. This approach that we call ``astrocladistics'' has been first applied to Dwarf Galaxies
of the Local Group and provides the first evolutionary galaxy tree. The cladogram
is sufficiently solid to support the existence of a hierarchical
organization in the diversity of galaxies, making it possible to track ancestral
types of galaxies. We also find that morphology is a summary of more fundamental
properties. Astrocladistics applied to cosmology simulated galaxies can, unsurprisingly, reconstruct 
the correct ``genealogy''. It reveals evolutionary lineages, divergences from common ancestors, character 
evolution behaviours and shows how mergers organize galaxy diversity. Application to real normal galaxies 
is in progress. Astrocladistics opens a new way to analyse galaxy evolution and a path towards a new 
systematics of galaxies.
\end{abstract}


\section{Introduction}
Galaxies have been found by Hubble to be isolated systems of stars (Hubble 1922, 1926). We know they also contain gas and dust, while black holes are certainly rather common. The Hubble (1936) diagram was originally devised to introduce evolution as a link between the different morphological types. This unification scheme has lost its meaning, but the Hubble tuning fork remains a reference in depicting galaxy diversity. However, numerous classifications, attached to particular observational criteria, have appeared, rendering difficult a visualisation of evolution of galaxies globally.
 
The biological world is also made of complex objects in evolution. The hierarchical organization of diversity on a tree-like structure is caused by evolution in a ``descent with modification'' scheme (Darwin 1859). A long history of classification of living organisms has lead to the establishment of powerful methodologies to reveal the evolutionary relationships between the different species. One of the most widely used nowadays is cladistics (Hennig 1965). It does not compare objects on appearance or global similarity, but rather on the evolutionary states of their characters (descriptors). 

An attempt to apply such a methodology to astrophysics was proposed by Fraix-Burnet et al (2003). Astrocladistics has thus been developed through some conceptualization and formalization of galaxy classification, formation, evolution and diversification. Several samples have now been analysed. In this talk, I present a summary of the present status of astrocladistics and its achievements. More details are to be found elsewhere (Fraix-Burnet et al 2004a, 2004b, 2004c, Fraix-Burnet and Davoust 2004). 

\section[]{Some basics of astrocladistics}

The detailed presentation of astrocladistics with its formalism, concepts and tools, is done in Fraix-Burnet et al (2004b, 2004c). Only a very brief overview is given in this section.

A galaxy is fully described and characterized by its basic constituents (stars, gas, dust) and all their properties. A galaxy is thus a complex object. Its evolution is the evolution of its constituents, so that their properties can be seen as characters with different evolutionary states. Hence a galaxy is represented by a vector having as many components as characters or descriptors allowed by observations. 

The formation of a galaxy should be understood as the formation of the galaxy as we see it, that is with all the character states as observed. Hence any physical or chemical process that affects any of its properties may result in the formation of a new galaxy of a different class or species. We have identified five such formation processes (Fraix-Burnet et al 2004c): assembling, secular evolution, interaction, accretion-merging, ejection-sweeping. In each case, material (basic constituents) of the progenitors is transmitted to the descendant while being modified. This is fundamentally the darwinian ``descent with modification'' mechanism. A big difference is that in the case of galaxies the modification is much more brutal because a galaxy can give birth to a new galaxy of a different species. 

Evolution of galaxies is governed by these formation processes which occur several times during the history of the Universe, always in different conditions. Diversity is thus generated, and is organized in a hierarchical way. Note that this has nothing to do with the hierarchical scenario of galaxy formation that considers only the mass. The astrocladistic analysis, using the matrix made with the galaxies and their characters, tries to arrange the objects on a tree-like structure, grouping them from the derived (evolved, that is inherited from a common ancestor) character states they share. It is then possible to define species or classes, and thus build a phylogenetic classification.

\section[]{Analyses of two samples of simulated galaxies}

In order to illustrate the validity of our approach and to better understand the application of cladistics to astrophysics, two samples of simulated galaxies have been analysed (Fraix-Burnet et al 2004b, 2004c). They were extracted from the
GALICS (Galaxies In Cosmological Simulations) database, which results from a hybrid model for hierarchical
galaxy formation studies, combining the outputs of large cosmological N-body simulations
with simple, semi-analytic recipes to describe the fate of the baryons within dark matter haloes (Hatton et al 2003). The main advantage is that the entire history is known for each galaxy with all the formation processes and progenitors. Note that cladistics does not aim at reconstructing genealogies (which in reality would require the impossible identification of all parents), but phylogenetic trees (which reveal the evolutionary relationships). Galaxies in GALICS are characterized principally by photometric properties (about 100).

The first sample (50 galaxies) is made with galaxies that have never known any merging, so that their diversification is due simply to secular evolution and gas accretion (no interaction are considered in the simulations). We show (Fraix-Burnet et al 2004b) that astrocladistics correctly establishes the right chronology. Diversity in this case is organized in a hierarchical pattern, as expected.

Among the five formation processes, merging of galaxies is somewhat peculiar. First, it clearly illustrates how galaxies disappear to form a new one, an important concept in astrocladistics, whatever the formation process. Second, the transmitted material of the merging galaxies is mixed together. This might look like biological hybridization which is known to affect the tree-like structure of diversity. However we show that this is not the case and confirm our conclusion using a sample with 43 galaxies having underwent different numbers of mergers in their history (Fraix-Burnet et al 2004c).

\section[]{Origin of the morphological dichotomy in the Dwarf galaxies of the Local Group}

The first application of astrocladistics to real galaxies has been performed on the Dwarf galaxies of the Local Group (Fraix-Burnet et al 2004a). The 36 objects are described by 24 characters (photometry, masses, line fluxes, metallicity, kinematics) plus morphology (spheroidal, irregular or intermediate). Since this last parameter is the only one to be qualitative and subjective, it has not been considered in the analysis but merely projected on the result tree. 

A robust tree is found, showing for the first time that diversity among a sample of real galaxies is organized hierarchically. Not only does this validates our approach, but the power of a cladistic analysis can now be exploited for galaxies. One very important result is that the morphological segregation is found {\it and} explained: irregular galaxies are all gathered on a branch which separated from spheroidal galaxies sometimes in the evolution because of a strong accretion of hydrogen. This is an indication that morphology is kind of a gross summary of more fundamental properties. However, the tree brings much more information in the form of consistent evolutionary interpretations. For instance, some objects are known to be exceptionally bright for spheroidal dwarf galaxies (as bright as irregulars). Our result yields a simple explanation: luminosity increases during evolution and these objects happen to be more evolved than other spheroidals and as much as irregulars. 

\section[]{Virgo galaxies: first step toward a new classification}

Two samples of galaxies belonging to the Virgo cluster are currently being analysed (Fraix-Burnet and Davoust 2004). Results are still preliminary, but a robust tree has been found with a sub-sample of 21 galaxies described by 35 characters (large band and line photometry, some kinematical data and line ratios). This shows that the hierarchical organization of galaxy diversity exists among real galaxies. Equally important, our study reveals a large consistency among evolution of most characters. For instance blue luminosity, B-V and U-B colours, line widths, maximum velocity dispersion, do all evolve quite regularly along the tree. It is thus possible to define different classes that are evolutionary related. More work has to be done for a complete interpretation. 

In the analysis, like in others, qualitative characters such as morphological type and presence/absence of a bar were not used. But their projection afterwards onto the result tree is very informative. First, the morphological type evolves quite regularly along the tree, again demonstrating that it is kind of a gross summary of more fundamental and quantitative properties of galaxies. Second, the character 'presence or absence of a bar' is erratic on the tree, being apparently not relevant for the evolutionary state of a galaxy. 

\section[]{Conclusions}

Astrocladistics is now producing more and more results while the concepts and tools have been largely clarified in the frame of galaxy diversity and evolution. The two fundamental ingredients (a galaxy is described by evolutive characters, diversity is caused by evolution and organized in a hierarchy) have been validated.

The application of cladistics to galaxy evolution requires a precise formalization of some definitions and concepts. In particular, it makes a difference between the definition, the description and the nomenclature of galaxies. This leads to the obvious result that the evolution of a galaxy is the evolution of its fundamental constituents. Their properties are the true descriptors of galaxies and characterize their evolution. This has been successfully checked on simulated and real galaxies.

The hierarchical organization of galaxy diversity, expected from the formation processes that generate new galaxy species, is found in all samples studied so far, either simulated or real. This is an extremely important result because if the existence of a tree-like structure remains true for all classes of galaxies at all epochs, it then will be possible to track the ancestral galaxy species back to the very first objects of the Universe.

Astrocladistics is a methodology to understand galaxy diversity via evolutionary relationships. It groups objects according to their history and using evolutionary states of their characters. It naturally builds a classification, but this can be done only after several analyses of different kinds of galaxies are achieved. This will be the next step of our work and this new galaxy classification will be based on solid taxonomical rules.


\end{document}